\RequirePackage{nag}
\documentclass[preprint,12pt,3p]{elsarticle}
\journal{Nuclear Physics B}
\usepackage{amssymb}
\usepackage[utf8]{inputenc}
\usepackage{graphicx}
\usepackage{color}
\usepackage{epstopdf}
\usepackage[nooneline,tight]{subfigure}
\newcommand{\spin}[1]{\sigma_{#1}} 
\newcommand{\nn}[2]{\left<#1,#2\right>} 
\newcommand{\nnn}[2]{\left<\left<#1,#2\right>\right>} 
\newcommand{\plaq}[4]{\left[#1,#2,#3,#4\right]} 
\newcommand{\nnsum}{\sum\limits_{\nn{i}{j}}\spin{i}\spin{j}} 
\newcommand{\nnnsum}{\sum\limits_{\nnn{i}{j}}\spin{i}\spin{j}} 
\newcommand{\plaqsum}{\sum\limits_{\plaq{i}{j}{k}{l}}%
  \spin{i}\spin{j}\spin{k}\spin{l}}
\def\be{\begin{equation}}
\def\ee{\end{equation}}
\def\bea{\begin{eqnarray}}
\def\eea{\end{eqnarray}}

\begin{document}
\title{Planar ordering in the plaquette-only gonihedric Ising model}
\author[itp]{Marco Mueller}
\ead{Marco.Mueller@itp.uni-leipzig.de}
\author[itp]{Wolfhard Janke}
\ead{Wolfhard.Janke@itp.uni-leipzig.de}
\author[hwu]{Desmond A. Johnston\corref{cor1}}
\cortext[cor1]{Corresponding author}
\ead{D.A.Johnston@hw.ac.uk}
\address[itp]{Institut f\"ur Theoretische Physik, Universit\"at Leipzig,\\ Postfach 100\,920, D-04009 Leipzig, Germany}
\address[hwu]{Department\ of Mathematics and the Maxwell Institute for Mathematical Sciences, Heriot-Watt University, Riccarton, Edinburgh, EH14 4AS, Scotland}
\begin{abstract}
In this paper we conduct a careful multicanonical simulation of the isotropic
$3d$ plaquette (``gonihedric'') Ising model and confirm that a planar,
fuki-nuke type order characterises the low-temperature phase of the model. From
consideration of the anisotropic limit of the model we define a class of order
parameters which can distinguish the low- and high-temperature phases in both
the anisotropic and isotropic cases. We also verify the recently voiced
suspicion  that the order parameter like behaviour of the standard magnetic
susceptibility $\chi_m$ seen in previous Metropolis simulations was an artefact
of the algorithm failing to explore the phase space of the macroscopically
degenerate low-temperature phase. $\chi_m$ is therefore not a suitable order
parameter for the model. 
\end{abstract}
\maketitle
\section{Introduction}
The  $3d$ plaquette (``gonihedric'') Ising Hamiltonian displays an unusual
planar flip symmetry, leading to an exponentially degenerate low-temperature
phase and non-standard scaling at its first-order phase transition point
\cite{goni_prl,goni_muca}. The nature of the order parameter for the plaquette
Hamiltonian has not been fully clarified, although simulations using a standard
Metropolis algorithm \cite{goni_order} have indicated that  the magnetic
ordering remains a fuki-nuke type planar layered ordering, which can be shown
rigorously to occur in the extreme anisotropic limit when the plaquette
coupling in one direction is set to zero \cite{suzuki_old,suzuki1,castelnovo} by mapping
the model onto a stack of $2d$ Ising models.

The  simple $3d$ plaquette Ising Hamiltonian, where the Ising spins $\sigma_i =
\pm 1$ reside on the vertices of a $3d$ cubic lattice,
\begin{equation}
\label{e2k}
H =  -  \frac{1}{2} \sum_{[i,j,k,l]}\sigma_{i} \sigma_{j}\sigma_{k} \sigma_{l}\;,
\end{equation}
can be considered as the $\kappa=0$ limit of a family of $3d$ ``gonihedric''
Ising Hamiltonians \cite{savvidy}, which contain nearest neighbour $\langle i,j
\rangle$, next-to-nearest neighbour $\langle \langle i,j \rangle \rangle$ and
plaquette interactions $[i,j,k,l]$,
\begin{equation}
  H^\kappa = -2\kappa\nnsum+\frac{\kappa}{2}\nnnsum-\frac{1-\kappa}{2}\plaqsum\;.
  \label{eq:ham:goni}
\end{equation} 
For $\kappa \ne 0$ parallel, non-intersecting planes of spins may be flipped in
the ground state at zero energy cost, leading to a $3 \times 2^{2L}$ 
ground-state degeneracy on an $L \times L \times L$ cubic lattice which is broken 
at finite temperature \cite{pietig_wegner}.  For the $\kappa=0$ plaquette
Hamiltonian of Eq.~(\ref{e2k}), on the other hand, the planar flip symmetry
persists throughout the low-temperature phase and extends to intersecting
planes of spins. This results in a macroscopic low-temperature phase degeneracy
of $2^{3L}$  and non-standard corrections to finite-size scaling at the
first-order transition displayed by the model \cite{goni_prl,goni_muca}.    

The planar flip symmetry of the low-temperature phase of the plaquette
Hamiltonian is intermediate between the global $\mathbb{Z}_2$ symmetry of  the
nearest-neighbour Ising model
\begin{equation}
\label{e0I}
  H_{\rm Ising} =  -  \sum_{\langle i,j \rangle}\sigma_{i} \sigma_{j}
\end{equation}
and the local gauge symmetry of a $\mathbb{Z}_2$ lattice gauge theory
\begin{equation}
  H_{\rm gauge} =  -  \sum_{[i,j,k,l]} U_{ij} U_{jk} U_{kl} U_{li}
  \label{e0G}
\end{equation}
and naturally poses the question of how to define a magnetic order parameter
that is sensitive to the first-order transition in the model. The standard
magnetization
\begin{equation}
  m = M/L^3 = \sum_{i} \sigma_i/L^3
\end{equation}
will clearly remain zero with periodic boundary conditions, even at lower
temperatures,  because of the freedom to flip arbitrary planes of spins.
Similarly, the absence of a local gauge-like symmetry means that observing the
behaviour of Wilson-loop type observables, as in a gauge theory, is also not
appropriate. 
\section{Fuki-Nuke Like Order Parameters}
Following the earlier work of Suzuki \emph{et~al.} \cite{suzuki_old,suzuki1}, a
suggestion for the correct choice of the order parameter for the isotropic
plaquette Hamiltonian comes from consideration of the $J_z=0$ limit of an {\it
anisotropic} plaquette model 
\bea
  H_{\rm aniso} &=& - J_x  \sum_{x=1}^{L} \sum\limits_{y=1}^{L}\sum\limits_{z=1}^{L} \sigma_{x,y,z} \sigma_{x,y+1,z}\sigma_{x,y+1,z+1} \sigma_{x,y,z+1} \nonumber \\
  &{}&  - J_y \sum_{x=1}^{L} \sum\limits_{y=1}^{L}\sum\limits_{z=1}^{L}  \sigma_{x,y,z} \sigma_{x+1,y,z}\sigma_{x+1,y,z+1} \sigma_{x,y,z+1}  \\
  &{}& - J_z \sum_{x=1}^{L} \sum\limits_{y=1}^{L}\sum\limits_{z=1}^{L}  \sigma_{x,y,z} \sigma_{x+1,y,z}\sigma_{x+1,y+1,z} \sigma_{x,y+1,z} \nonumber \;,
\eea
where we now indicate each site and directional sum explicitly, assuming we are
on a cubic $L \times L \times L$ lattice with periodic boundary conditions
$\sigma_{L+1, y, z} =
\sigma_{1,y,z}$, $\sigma_{x, L+1, z} = \sigma_{x,1,z}$, $\sigma_{x,y,L+1} =
\sigma_{x,y,1}$.
This will prove to be convenient in the sequel when discussing candidate order
parameters.

When $J_z=0$ the horizontal, ``ceiling'' plaquettes have zero coupling, which
Hashizume and Suzuki denoted the  ``fuki-nuke'' (``no-ceiling'' in Japanese)
model~\cite{suzuki1}. The anisotropic $3d$ plaquette Hamiltonian at $J_z=0$,
\bea
H_{\rm fuki-nuke} &=& - J_x  \sum_{x=1}^{L} \sum\limits_{y=1}^{L}\sum\limits_{z=1}^{L} \sigma_{x,y,z} \sigma_{x,y+1,z}\sigma_{x,y+1,z+1} \sigma_{x,y,z+1} \nonumber \\
&{}&  - J_y \sum_{x=1}^{L} \sum\limits_{y=1}^{L}\sum\limits_{z=1}^{L}  \sigma_{x,y,z} \sigma_{x+1,y,z}\sigma_{x+1,y,z+1} \sigma_{x,y,z+1}\;,
\eea
may be rewritten as a stack of $2d$ nearest-neighbour Ising models by defining
bond spin variables $\tau_{x,y,z} = \sigma_{x,y,z} \sigma_{x,y,z+1}$ at each
vertical lattice bond which satisfy $\prod_{z=1}^L\tau_{x,y,z} = 1$ trivially
for periodic boundary conditions.
The $\tau$ and $\sigma$ spins are related by
an inverse relation of the form
\be 
\sigma_{x,y,z} =\sigma_{x,y,1}\times \tau_{x,y,1} \, \tau_{x,y,2} \, \tau_{x,y,3} \cdots \tau_{x,y,z-1}\;,
\ee
and the partition function acquires an additional factor of $2^{L\times L}$ arising from the transformation.
%
%
The resulting Hamiltonian with $J_x = J_y = 1$ is then
simply that of a stack of decoupled $2d$ Ising layers with the
standard nearest-neighbour in-layer interactions in the horizontal planes,
\be
H_{\rm fuki-nuke} = - \sum\limits_{x=1}^{L}\sum\limits_{y=1}^{L}\sum\limits_{z=1}^{L} \left( \tau_{x,y,z}  \tau_{x+1,y,z} + \tau_{x,y,z} \tau_{x,y+1,z} \right)\;,
\label{stack}
\ee
apart from the $L^2$ constraints $\prod_{z=1}^L\tau_{x,y,z} = 1$ whose 
contribution should vanish in the thermodynamic limit~\cite{castelnovo}.
%
Each $2d$ Ising layer in Eq.~(\ref{stack}) will magnetize independently at the $2d$
Ising model transition temperature. A suitable order parameter in a single
layer is the standard Ising magnetization 
\be
  m_{2d, z} =  \left< \frac{1}{L^2} \sum_{x=1}^{L} \sum\limits_{y=1}^{L} \tau_{x,y,z} \right>
\label{Mone}
\ee
which when translated back to the original $\sigma$ spins gives
\be 
  m_{2d, z} = \left<  \frac{1}{L^2} \sum_{x=1}^{L} \sum\limits_{y=1}^{L} \sigma_{x,y,z} \sigma_{x,y,z+1}   \right> 
\ee
which will behave as $\pm \left| \beta - \beta_c \right|^{1 \over 8}$ near the
critical point $\beta_c = \frac{1}{2}\ln ( 1 + \sqrt{2})$.  More generally,
since the different $\tau_{x,y,z}$ layers are decoupled in the vertical
direction we could define
\be
m_{2d, \, z, \, n} = \left<\frac{1}{L^2} \sum_{x=1}^{L} \sum\limits_{y=1}^{L} \sigma_{x,y,z} \sigma_{x,y,z+n} \right> =   ( m_{2d, \, z} )^n \; .
\ee
Two possible options for constructing a pseudo-$3d$ order parameter suggest
themselves in the fuki-nuke case. One is to take the absolute value of the
magnetization in each independent layer
\be
m_{\rm abs} = \left< \frac{1}{L^3}\sum_{z=1}^{L} \left| \sum\limits_{x=1}^{L}\sum\limits_{y=1}^{L} \sigma_{x,y,z}\sigma_{x,y,z+1}\right| \right>\;,
\label{Mabs}
\ee
the other is to square the magnetization of each plane,
\be
  m_{\rm sq} =    \left< \frac{1}{L^5} \sum_{z=1}^{L}  \left( \sum\limits_{x=1}^{L}\sum\limits_{y=1}^{L}   \sigma_{x,y,z}\sigma_{x,y,z+1}  \right)^2  \right> \;,
\label{Msq}
\ee
to avoid inter-plane cancellations when the contributions from each Ising layer
are summed up.  We have explicitly retained the various normalizing factors in
Eqs.~(\ref{Mabs}) and (\ref{Msq}) for a cubic  lattice with $L^3$ sites. 

The suggestion in \cite{goni_order,suzuki1} was that similar order parameters could still
be viable for the isotropic plaquette action, namely
\begin{eqnarray}
  m_{\rm abs}^x = \left< \frac{1}{L^3}\sum_{x=1}^{L} \left|\sum\limits_{y=1}^{L}\sum\limits_{z=1}^{L} \sigma_{x,y,z}\sigma_{x+1,y,z}\right| \right>\;,\nonumber\\
  m_{\rm abs}^y = \left< \frac{1}{L^3}\sum_{y=1}^{L} \left|\sum\limits_{x=1}^{L}\sum\limits_{z=1}^{L} \sigma_{x,y,z}\sigma_{x,y+1,z}\right| \right>\;,\\
  m_{\rm abs}^z = \left< \frac{1}{L^3}\sum_{z=1}^{L} \left|\sum\limits_{x=1}^{L}\sum\limits_{y=1}^{L} \sigma_{x,y,z}\sigma_{x,y,z+1}\right| \right>\;, \nonumber
\end{eqnarray}
where we again assume periodic boundary conditions.  Similarly, for the case of the squared magnetizations one can
define 
\begin{eqnarray}
  m_{\rm sq}^x = \left< \frac{1}{L^5}\sum_{x=1}^{L} \left(\sum\limits_{y=1}^{L}\sum\limits_{z=1}^{L} \sigma_{x,y,z}\sigma_{x+1,y,z}\right)^2 \right>\;, \\
  \nonumber
\end{eqnarray}
with obvious analogous definitions for the other directions, $m_{\rm sq}^y$ and
$m_{\rm sq}^z$, which also appear to be viable candidate order parameters. In
the isotropic case the system should be agnostic to the direction so we would
expect $m_{\rm abs}^x = m_{\rm abs}^y = m_{\rm abs}^z$ and similarly for the
squared quantities.  The possibility of using an order parameter akin to the
$m_{\rm sq}$ in Eq.~(\ref{Msq}) had also been  suggested by Lipowski \cite{3a},
who confirmed that it appeared to possess the correct behaviour in a small
simulation.

In Ref.~\cite{goni_order} Metropolis simulations gave a strong indication that
$m_{\rm abs}^{x,y,z}$ and $m_{\rm sq}^{x,y,z}$ as defined above were indeed suitable
order parameters for the isotropic plaquette model, but these were subject to
the difficulties of simulating a strong first-order phase transition with such
techniques and also produced the possibly spurious result that the standard
magnetic susceptibility $\chi$  behaved like an order parameter. The
aforementioned difficulties precluded a serious scaling analysis of the
behaviour of the order parameter with Metropolis simulations, including an
accurate estimation of the transition point via this route.

With the use of the multicanonical Monte Carlo algorithm~\cite{muca,
janke_muca} in which rare states lying between the disordered and ordered
phases in the energy histogram are promoted artificially to decrease
autocorrelation times and allow more rapid oscillations between ordered and
disordered phases, combined with reweighting techniques~\cite{reweighting}, we
are able to carry out much more accurate measurements of $m_{\rm abs}^{x,y,z}$
and $m_{\rm sq}^{x,y,z}$. This allows us to confirm the suitability of the
proposed order parameters and to examine their scaling properties near the
first-order transition point. 

The results presented here can also be regarded as the magnetic counterpart of
the high-accuracy investigation of the scaling of energetic quantities (such as
the energy, specific heat and Binder's energetic parameter) for the
plaquette-only gonihedric Ising model and its dual carried out in
\cite{goni_muca} and provide further confirmation of the estimates of the
critical temperature determined there, along with the observed non-standard
finite-size scaling. 
\section{Simulation Results}
We now discuss in detail our measurements of the  proposed fuki-nuke
observables~\cite{goni_order} defined above, using  multicanonical simulation
techniques. The algorithm used is a two-step process, where we iteratively
improve guesses to an a priori unknown weight function $W(E)$ for
configurations $\left\{\sigma\right\}$ with system energy
$E=H(\left\{\sigma\right\})$ which replaces the Boltzmann weights
$e^{-\beta E}$ 
in the acceptance rate of traditional Metropolis Monte
Carlo simulations. In the first step the weights are adjusted so that the
transition probabilities  between configurations with different energies become
constant, giving a flat energy histogram \cite{mucaweights}. The second step is
the actual production run using the fixed weights produced iteratively in step
one. This yields the time series of the energy, magnetization and the two
different fuki-nuke observables $m_{\rm sq}$ and $m_{\rm abs}$ in their three
different spatial orientations. With sufficient statistics such time series
together with the weights can provide the 8-dimensional density of states
$\Omega(E, m, m_{\rm abs}^x, \dots)$, or by taking the logarithm the
coarse-grained free-energy landscape, by simply counting the occurrences of $E,
m, m_{\rm abs}^x, \dots$ and weighting them with the inverse $W^{-1}(E)$ of the
weights fixed prior to the production run. Practically, estimators of the
microcanonical expectation values of observables are used, where higher
dimensions are integrated over in favour of reducing the amount of statistics
required.

Although the actual production run consisted of $N=(100-1000)\times 10^6$ sweeps
depending on the lattice sizes and is therefore quite long, the statistics for
the fuki-nuke order parameters is sparser. For these, we carried out
measurements every $V=L^3$ sweeps, because one has to traverse the lattice once
to measure the order parameters in all spatial orientations and this has a
considerable impact on simulation times. With skipping intermediate sweeps we
end up with less statistics, but the resulting measurements are less correlated
in the final time series. More details of the statistics of the simulations are
given in~\cite{goni_muca}. 
\begin{figure}
  \begin{center}
    \includegraphics[width=0.5\textwidth]{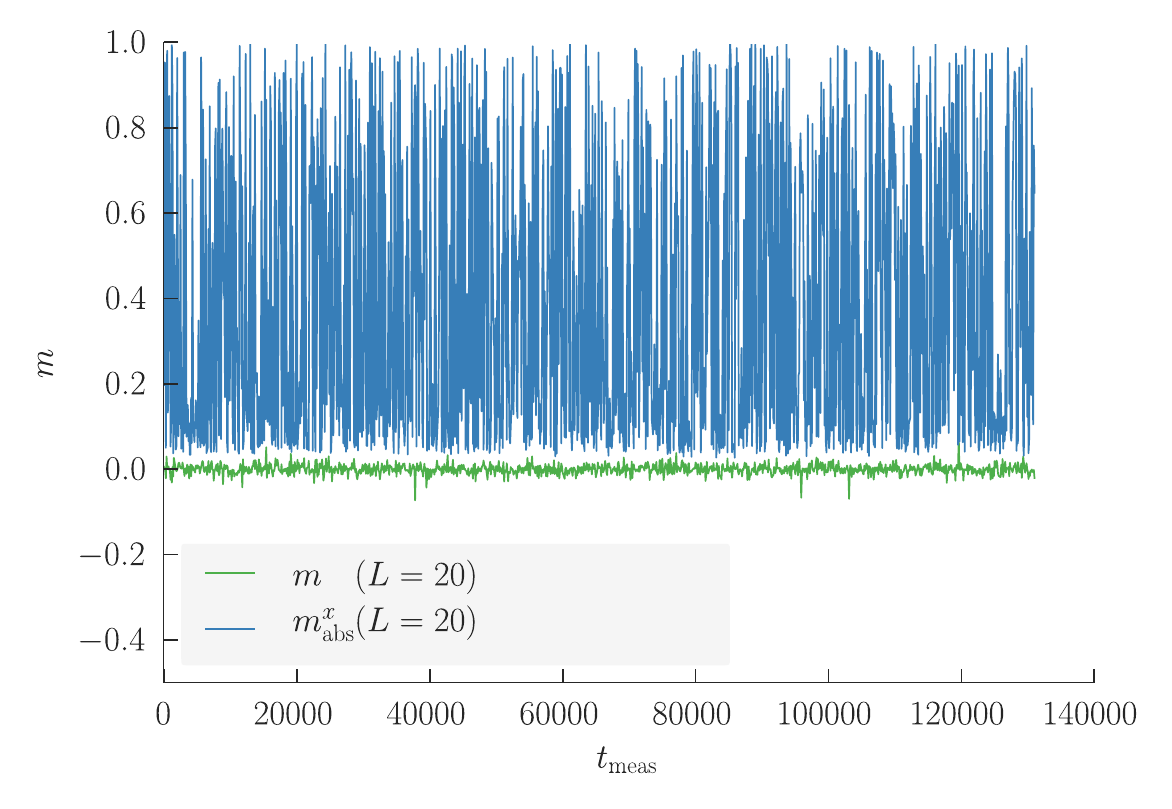}\hfill
    \includegraphics[width=0.5\textwidth]{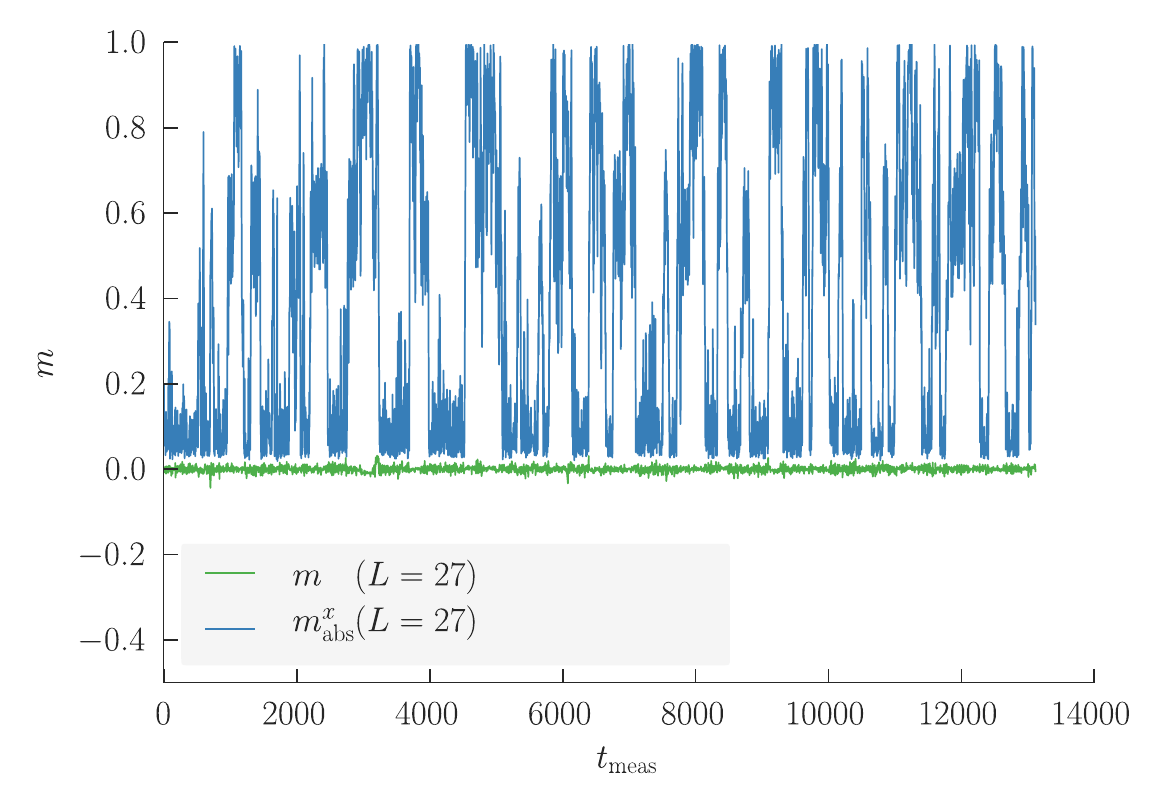}\\
    \subfigure[]{\includegraphics[width=0.5\textwidth]{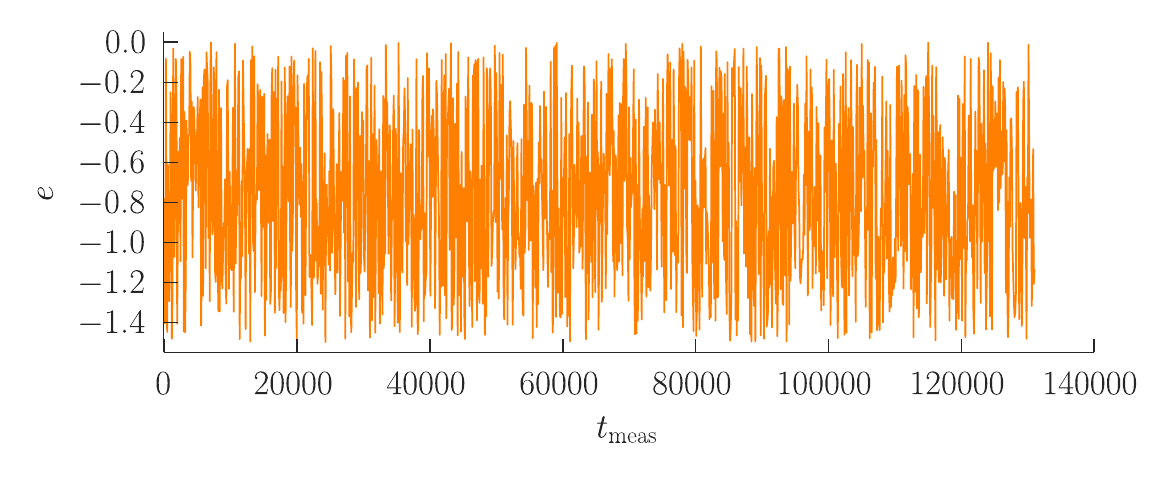}}\hfill 
    \subfigure[]{\includegraphics[width=0.5\textwidth]{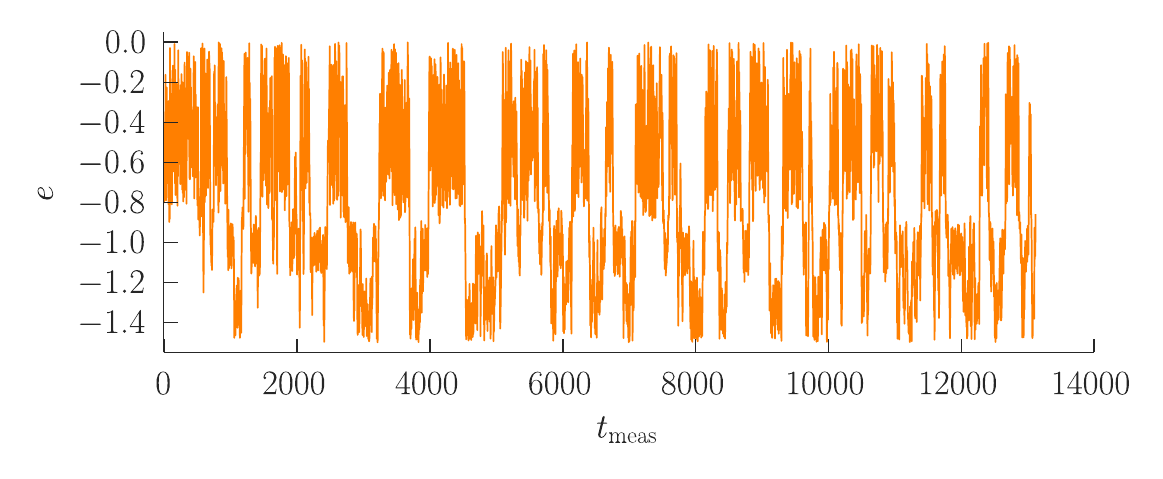}}
    \caption{(a) Example time series of the multicanonical simulations for an
      intermediate lattice with linear size $L=20$. The upper row shows the
      normalized magnetization $m$ and the fuki-nuke observable $m_{\rm
      abs}^x$, the lower row shows the energy per system volume 
      $e=E/L^3$. 
      (b) The same for a large lattice with linear size $L=27$. }
    \label{fig:timeseries} 
  \end{center} 
\end{figure} 
\begin{figure}
  \begin{center} 
    \includegraphics[width=0.45\textwidth]{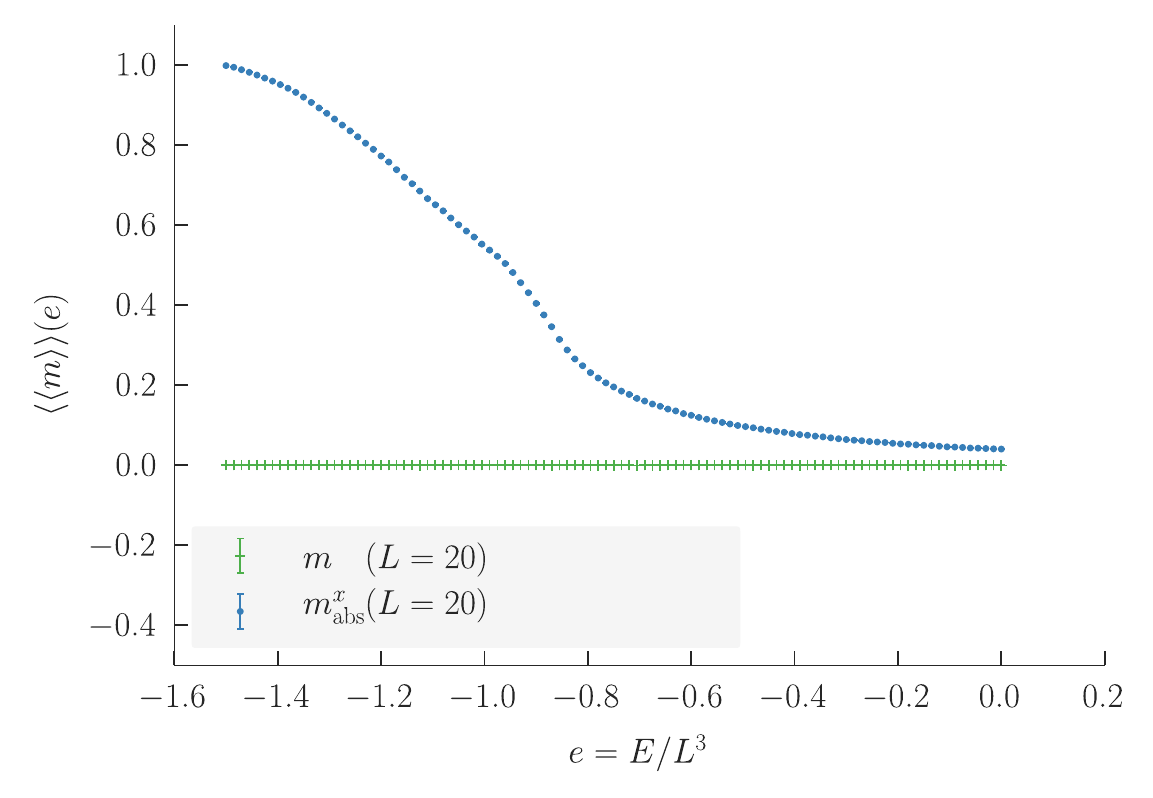}
    \includegraphics[width=0.45\textwidth]{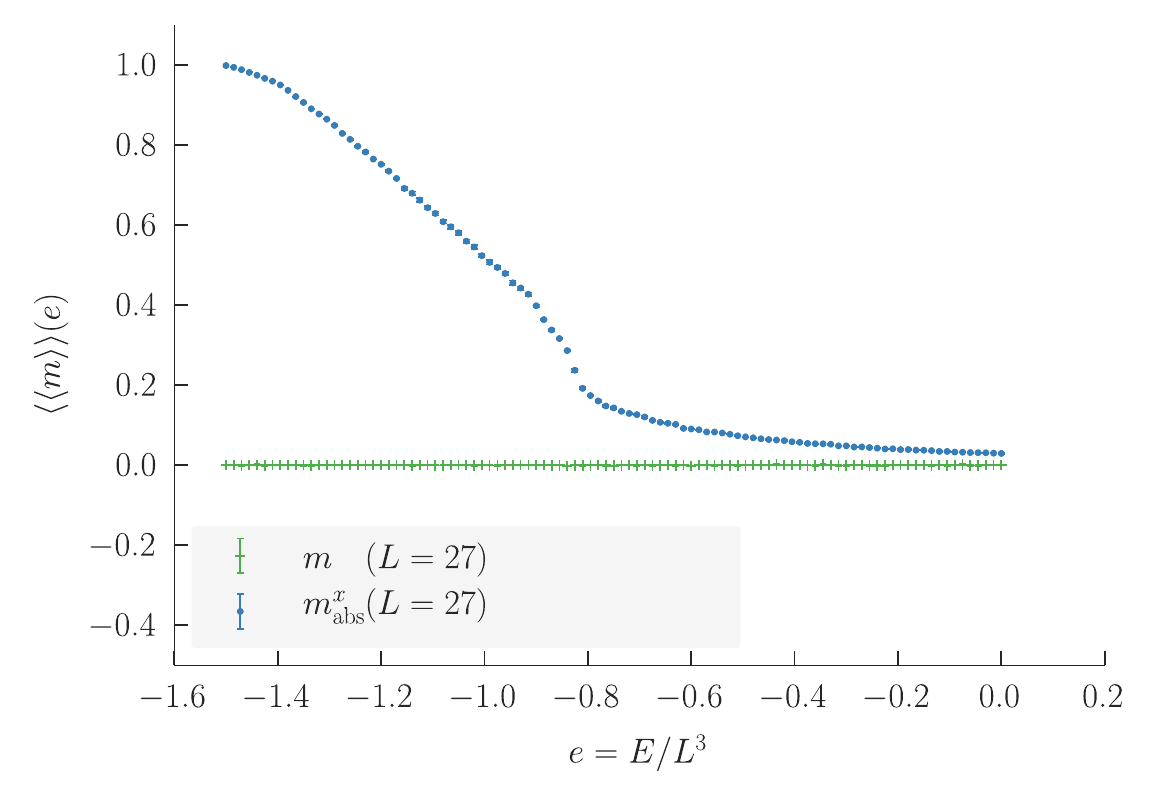}
    \caption{Microcanonical estimators for the magnetization $m$ and fuki-nuke
      parameter $m_{\rm abs}^x$ for lattices with size $L=20$ and $L=27$, where
      we used $100$ bins for the energy $e$ in this representation.
      Statistical errors are obtained from Jackknife error analysis with $20$
      blocks.}
    \label{fig:mE} 
  \end{center} 
\end{figure} 

\begin{figure}[t] 
  \begin{center} 
    \subfigure[]{\includegraphics[width=0.45\textwidth]{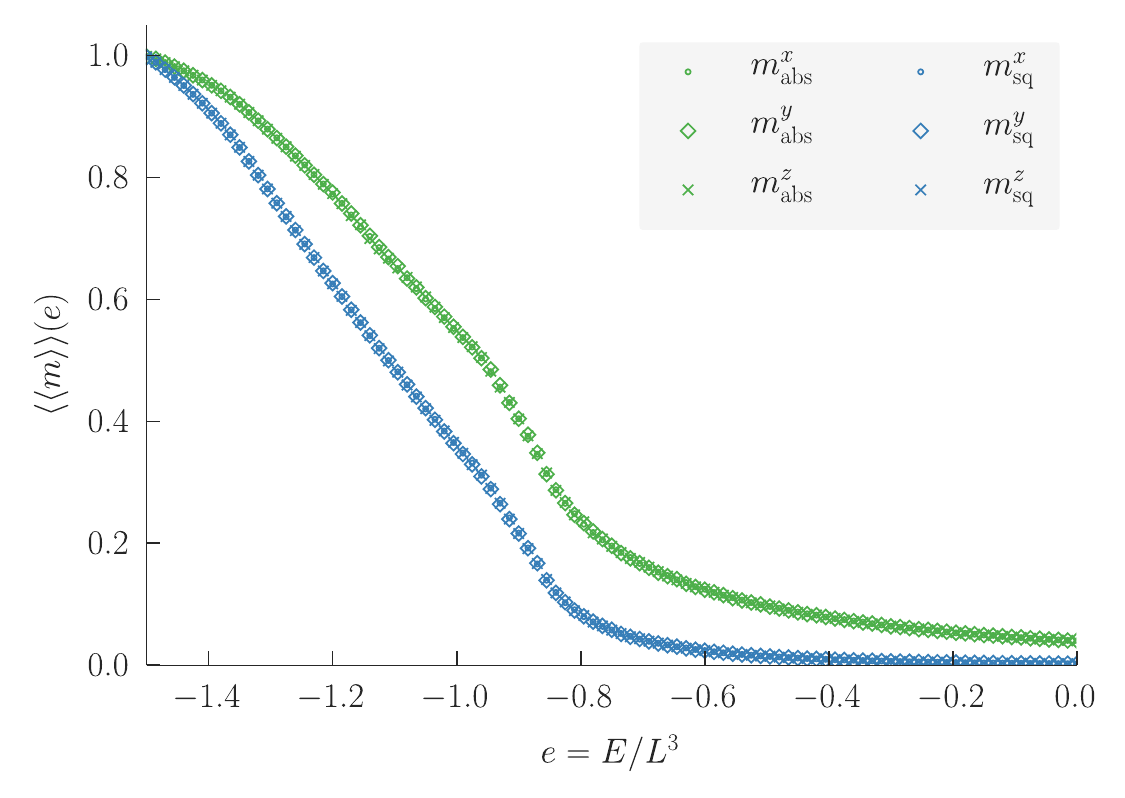}}
    \subfigure[]{\includegraphics[width=0.45\textwidth]{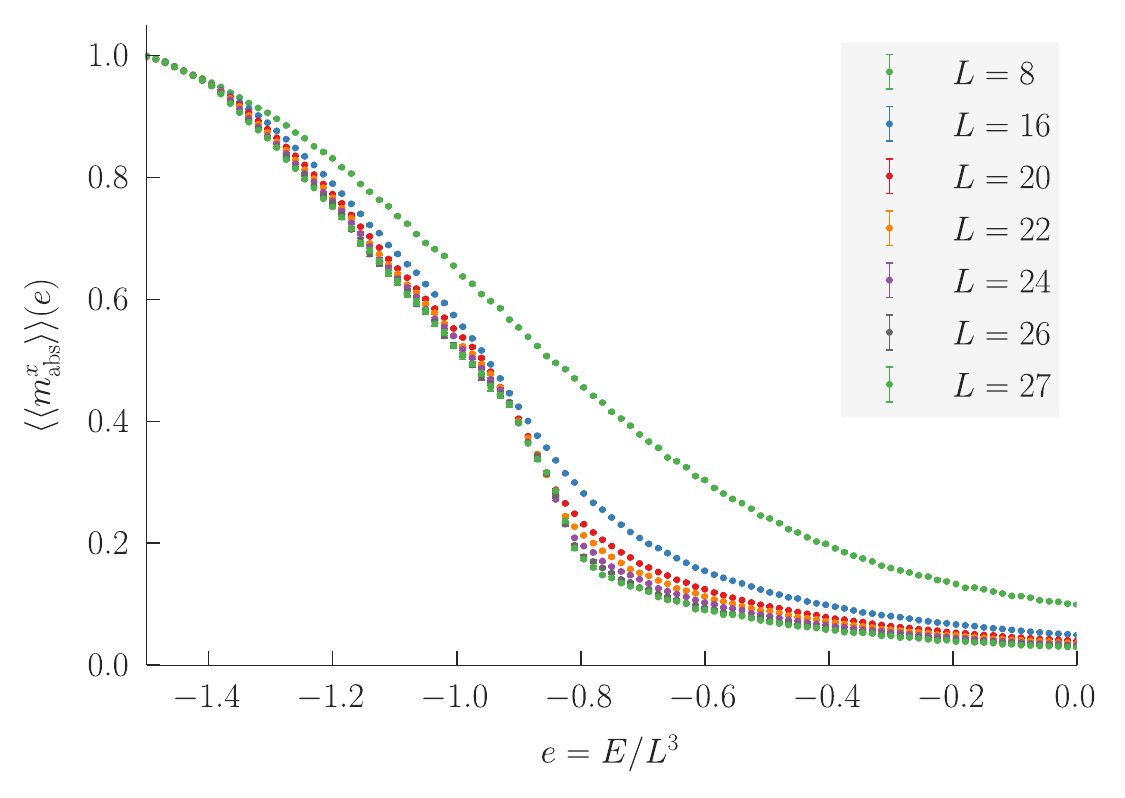}}
    \caption{(a) Microcanonical estimators for the different orientations of
      the fuki-nuke parameters $m^{x,y,z}_{\rm abs, sq}$ for a lattice with
      linear size $L=20$, which fall onto two curves. The statistical errors 
      are smaller than the data symbols and have been omitted for clarity.  
      (b) Microcanonical estimators for the fuki-nuke parameter $m^{x}_{\rm abs}$ 
      for several lattice sizes.}
  \label{fig:microequal} 
  \end{center} 
\end{figure}

In Fig.~\ref{fig:timeseries} we show the full time series for the magnetization
$m$ and the fuki-nuke parameter $m^x_{\rm abs}$ of the multicanonical
measurements for an intermediate {($L=20$)} and the largest {($L=27$)} lattice
size in the simulations along with the system energy. The time series of the
energy per system volume $e=E/L^3$ of the larger lattice can be
seen to be reflected numerous times at $e\simeq -0.9$ (coming from the
disordered phase) and $e\simeq -1.2$ (coming from the ordered phase) which
shows qualitatively that additional, athermal and non-trivial barriers may be
apparent in the system \cite{droplet}.  {It is clear} that the standard
magnetization is not a suitable order parameter, since it continues to
fluctuate around zero even though the system transits many times between
ordered and disordered phases {in the course of the simulation}.  The fuki-nuke
order parameter, on the other hand, shows a clear signal {for the transition},
tracking the jumps which are  visible in the energy time series.  This is also
reflected in Fig.~\ref{fig:mE}, where we show the estimators of the
microcanonical expectation values $\langle\langle \cdot \rangle\rangle$ of our
observables~$O$,
\begin{eqnarray}
  \langle\langle O \rangle\rangle(E) = \sum\limits_O O\, \Omega(E,O) \Big/ \sum\limits_O\Omega(E,O)\;,
  \label{eq:micro-magnetic}
\end{eqnarray}
where the quantity $\Omega(E,O)$ is the number of states with energy $E$ and
value $O$ of any of the observables, in this case either the magnetization $m$
or one of the fuki-nuke parameters.  We get an estimator for $\Omega(E,O)$ by
counting the occurrences of the pairs $(E,O)$ in the time series and weighting
them with $W^{-1}(E)$. For clarity in the graphical representation in
\mbox{Figs.~\ref{fig:mE} and \ref{fig:microequal}} we only used a partition of
$100$ bins for the energy interval. 
An estimate for the statistical error of each bin is calculated by Jackknife
error analysis~\cite{jackknife}, decomposing the time series into $N_{\rm B}=20$
non-overlapping blocks of length $b=N/N_{\rm B}$.
The $i$-th Jackknife estimator is given by 

\begin{eqnarray}
  \langle\langle O \rangle\rangle^i(E) = \sum\limits_O O\, \Omega^i(E,O) \Big/ \sum\limits_O\Omega^i(E,O)\;,
  \label{eq:jkMicro}
\end{eqnarray}
with $\Omega^i$ being the occurrences of pairs $(E,O)$ in a reduced time 
series where the $i$-th block of length $b$ has been omitted.
The variance of the Jackknife estimators then is proportional to the
squared statistical error of their mean $\langle\overline{\langle O
\rangle\rangle^i(E})=(1/N_{\rm B})\sum_{i=1}^{N_{\rm B}}\langle\langle O
\rangle\rangle^i(E)\approx\langle\langle O \rangle\rangle(E)$, \begin{eqnarray}
  \varepsilon^2_{\langle\overline{\langle O \rangle\rangle^i(E})} =
  \frac{N_{\rm B} - 1}{N_{\rm B}} \sum\limits_{i=1}^{N_{\rm B}} \left(
  \langle\langle O \rangle\rangle^i(E)- \langle\overline{\langle O
  \rangle\rangle^i(E}) \right)^2\;, \label{eq:jkerr} \end{eqnarray} where the
prefactor $(N_{\rm B} -1)/N_{\rm B}$ accounts for the trivial correlations
caused by reusing each data point in $N_{\rm B} - 1$
estimators~\cite{jackknife}.  Aside from reducing a systematic bias in derived
quantities, the Jackknife error  automatically takes care of temporal
correlations as long as the block length $b$ is greater than the
autocorrelation time which was measured and discussed in great detail
in~\cite{goni_muca}.  We additionally confirmed that $N_{\rm B} = 10$ and
$N_{\rm B}=40$ for selected lattice sizes yield the same magnitude for the
statistical error.

That the fuki-nuke parameters are, indeed,  capable of distinguishing ordered
and disordered states is depicted in \mbox{Figs.~\ref{fig:timeseries} and
\ref{fig:mE}}.  In the microcanonical picture we can clearly see that the
different orientations of the fuki-nuke parameters are equal for the isotropic
gonihedric Ising model, which we show for $L=20$ in
Fig.~\ref{fig:microequal}(a). This confirms that the sampling is at least
consistent in the simulation.  We also collect the microcanonical estimators
for $m^x_{\rm abs}$ for several lattice sizes in Fig.~\ref{fig:microequal}(b),
where a region of {approximately} linear increase between  $e\simeq-0.9$ and
$e\simeq-1.3$ can be seen for the larger lattices.  This interval corresponds
to the energies of the transitional, unlikely states between the ordered and
disordered phases.  Plaquettes successively become satisfied towards the
ordered phase and thus the estimators that measure intra- and inter-planar
correlations must increase, too.
\begin{figure} 
  \begin{center} 
    \includegraphics[width=0.45\textwidth]{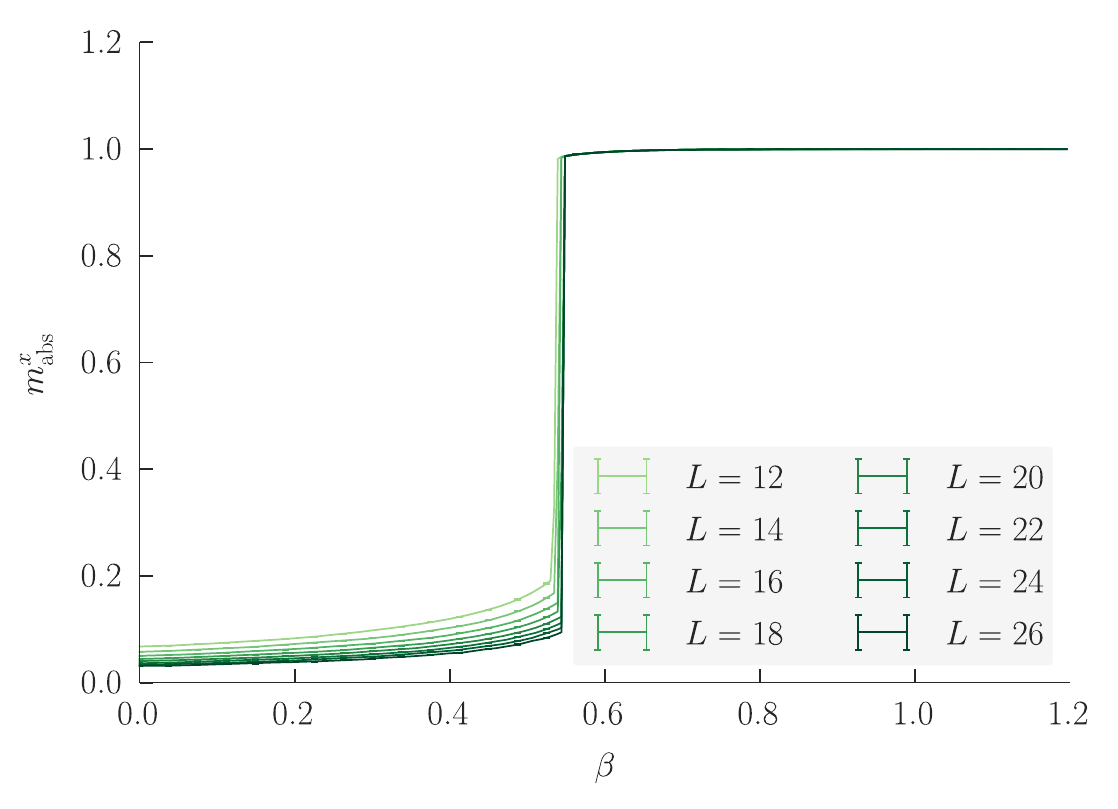}
    \includegraphics[width=0.45\textwidth]{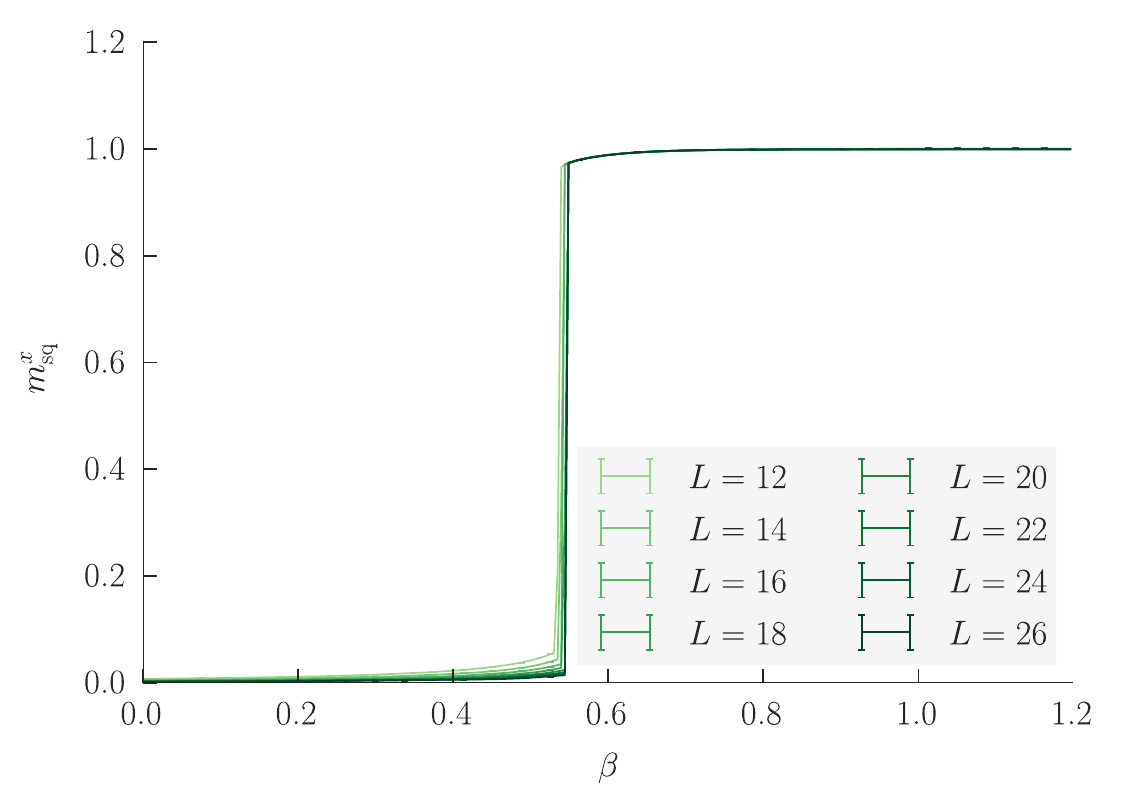}
    \caption{Canonical curves for the fuki-nuke parameters $m^{x}_{\rm abs}$
      and $m^{x}_{\rm sq}$ over a broad range of inverse temperature $\beta$ for
      several lattice sizes $L$ (compare with the canonical data
      in~\cite{goni_order}).}
  \label{fig:cmp2012} 
  \end{center} 
\end{figure} 

As we stored the full time series along with its weight function, we are able
to measure the microcanonical estimators for arbitrary functions of the
measured observables $f(O)$,
\begin{eqnarray}
  \langle\langle f(O) \rangle\rangle(E) = \sum\limits_O f(O)\, \Omega(E,O) \Big/ \sum\limits_O\Omega(E,O)\;,
  \label{eq:micro}
\end{eqnarray}
which can be exploited {to give} a convenient way of calculating higher-order
moments as well. For canonical simulations reweighting
techniques~\cite{reweighting} allow the inference of system properties in a
narrow range around the simulation temperature. That range and the accuracy are
then determined by the available statistics of the typical configurations for
the temperature of interest.  Since multicanonical simulations yield histograms
with statistics covering a broad range of energies, which is {their} most
appealing feature and common to flat-histogram techniques, it is possible to
reweight to a broad range of temperatures.  The canonical estimator at
finite inverse temperature $\beta>0$ is thus obtained by
\begin{eqnarray}
  \langle O\rangle(\beta) = \sum\limits_E \langle\langle O\rangle\rangle (E) \, 
  e^{-\beta E} \Big/ \sum\limits_E e^{-\beta E}\;,
  \label{eq:cano}
\end{eqnarray}
and again Jackknife error analysis is employed for an estimate of the
statistical error, where we form individual Jackknife
estimators by inserting the $i$-th microcanonical estimator of
Eq.~(\ref{eq:jkMicro}) into Eq.~(\ref{eq:cano}) and apply an analogue to
Eq.~(\ref{eq:jkerr}).

Since the microcanonical estimators for different
orientations agree within error bars the canonical values will also be the
same. Therefore, in Fig.~\ref{fig:cmp2012} we only show one orientation for the
two different fuki-nuke parameters. 
The overall behaviour of $m^{x}_{\rm abs}$ and $m^{x}_{\rm sq}$ from the
Metropolis simulations of Ref.~\cite{goni_order} is recaptured by the
multicanonical data here. Namely, sharp jumps are found near the inverse
transition temperature,  as expected for an order parameter at a first-order
phase transition. The transition temperature that was determined in the
earlier simulations where energetic observables were measured under different
boundary conditions and from a duality relation was 
$\beta = 0.551\,334(8)$~\cite{goni_muca}. The positions {of the jumps seen here 
(and in the earlier
simulations)} depend on the lattice size, and finite-size scaling can be
applied to {estimate} the transition temperature under the assumption that the
fuki-nuke parameters are indeed {suitable} order parameters.

\begin{figure} 
  \begin{center} 
    \includegraphics[width=0.45\textwidth]{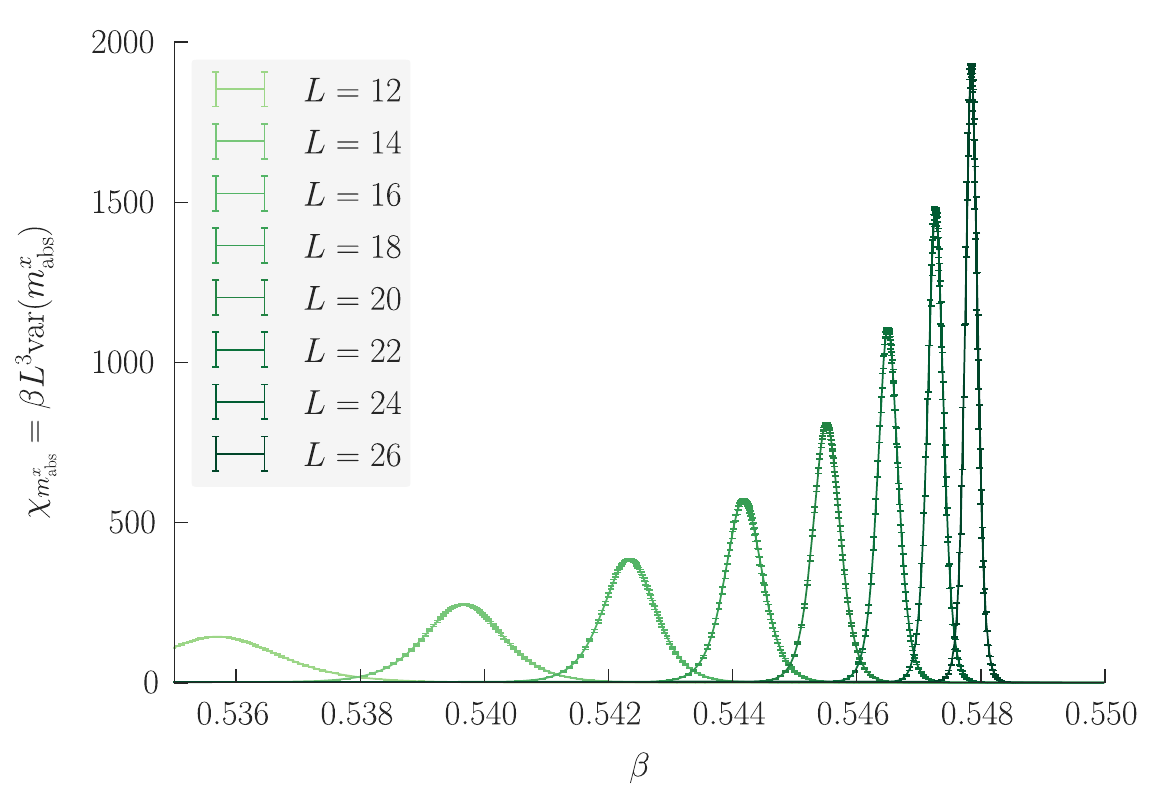}
    \includegraphics[width=0.45\textwidth]{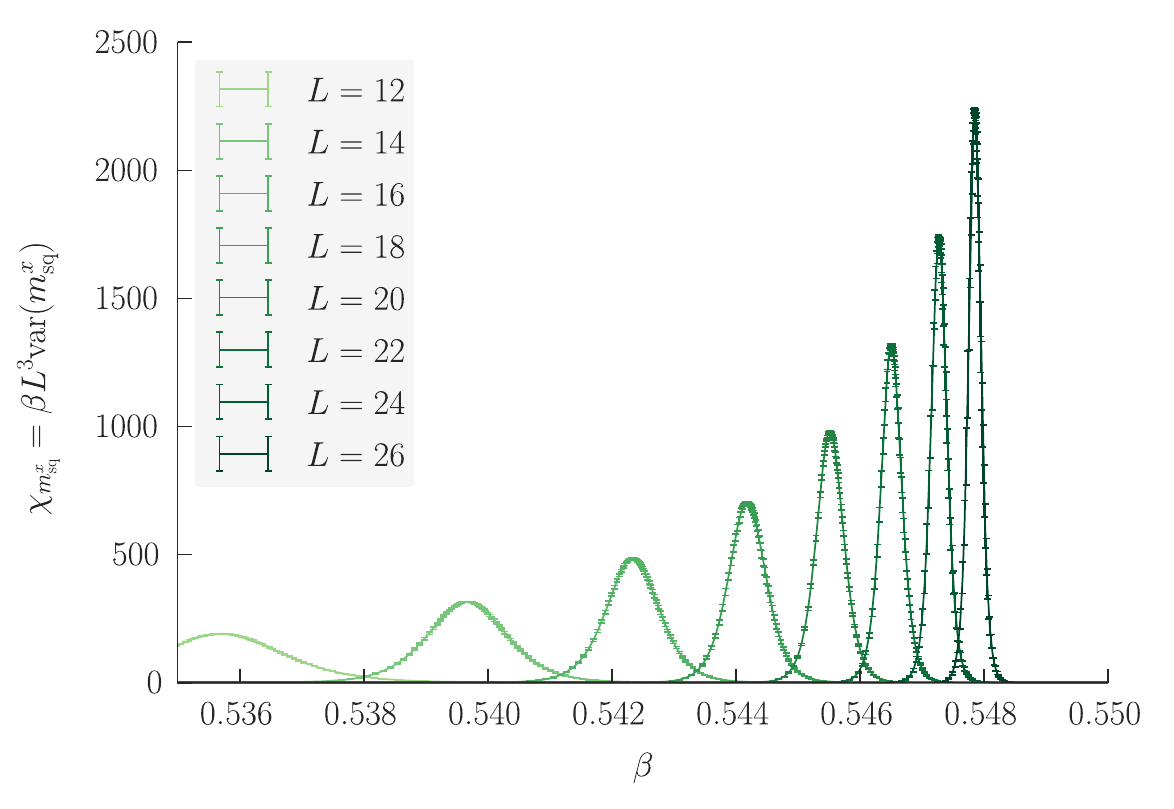}
    \caption{Canonical curves for the susceptibilities of fuki-nuke parameters
      $m^{x}_{\rm abs}$ and $m^{x}_{\rm sq}$ near the phase transition
      temperature for different lattice sizes.}
    \label{fig:canonchi} 
  \end{center} 
\end{figure}

To carry out such a finite-size scaling analysis, it is advantageous to look at
the canonical curves of the susceptibilities, \mbox{$\chi_O(\beta) = \beta L^3
\left( \langle O^2 \rangle(\beta) - \langle O \rangle(\beta)^2 \right)$},
{since their} peak positions provide an accurate measure of the finite-lattice
inverse transition temperature $\beta^{\chi_O}(L)$. {As an example}
Fig.~\ref{fig:canonchi} {shows} the peaks of the susceptibilities belonging to
$m_{\rm abs}^x$ and $m_{\rm sq}^x$ for several lattice sizes. Qualitatively
{the behaviour of} both susceptibilities {is similar} and 
as for the specific heat~\cite{goni_muca,procedia} 
their maxima scale proportional to the system volume $L^3$ but they differ in 
their magnitudes.
\begin{figure}[t]
  \begin{center} 
    \includegraphics[width=0.5\textwidth]{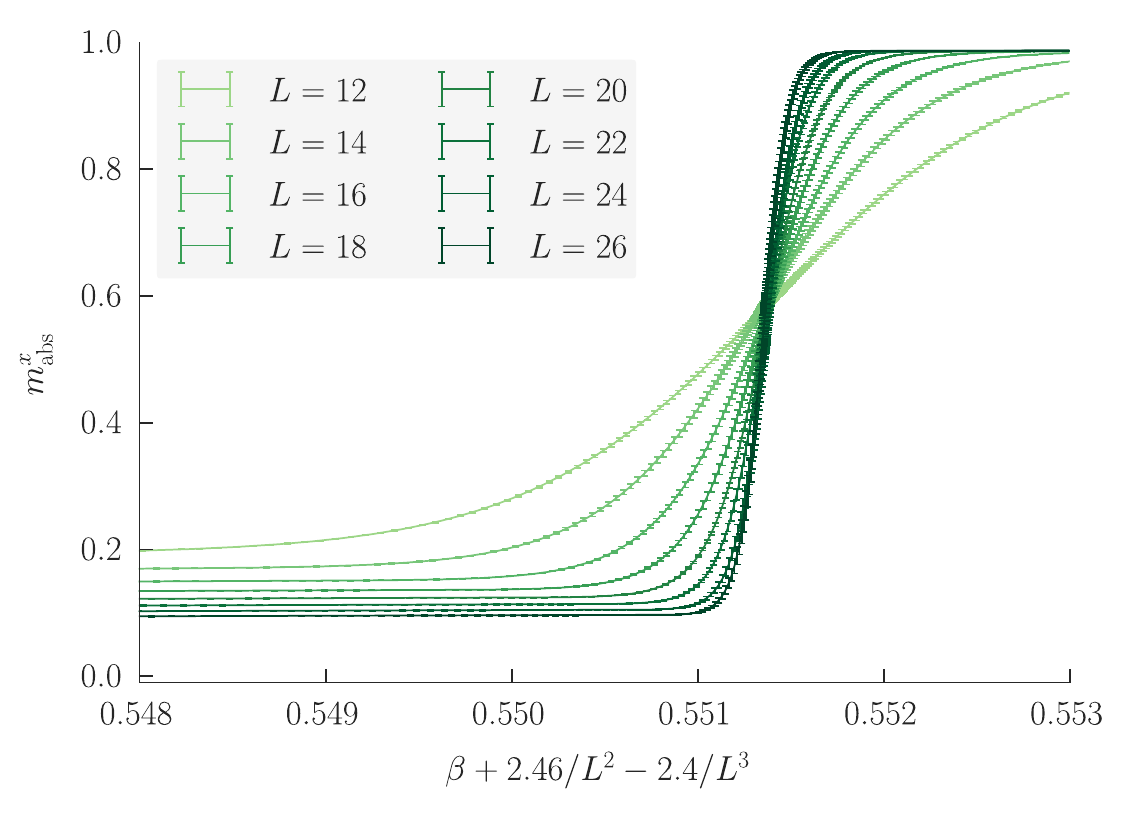}\hfill
    \includegraphics[width=0.5\textwidth]{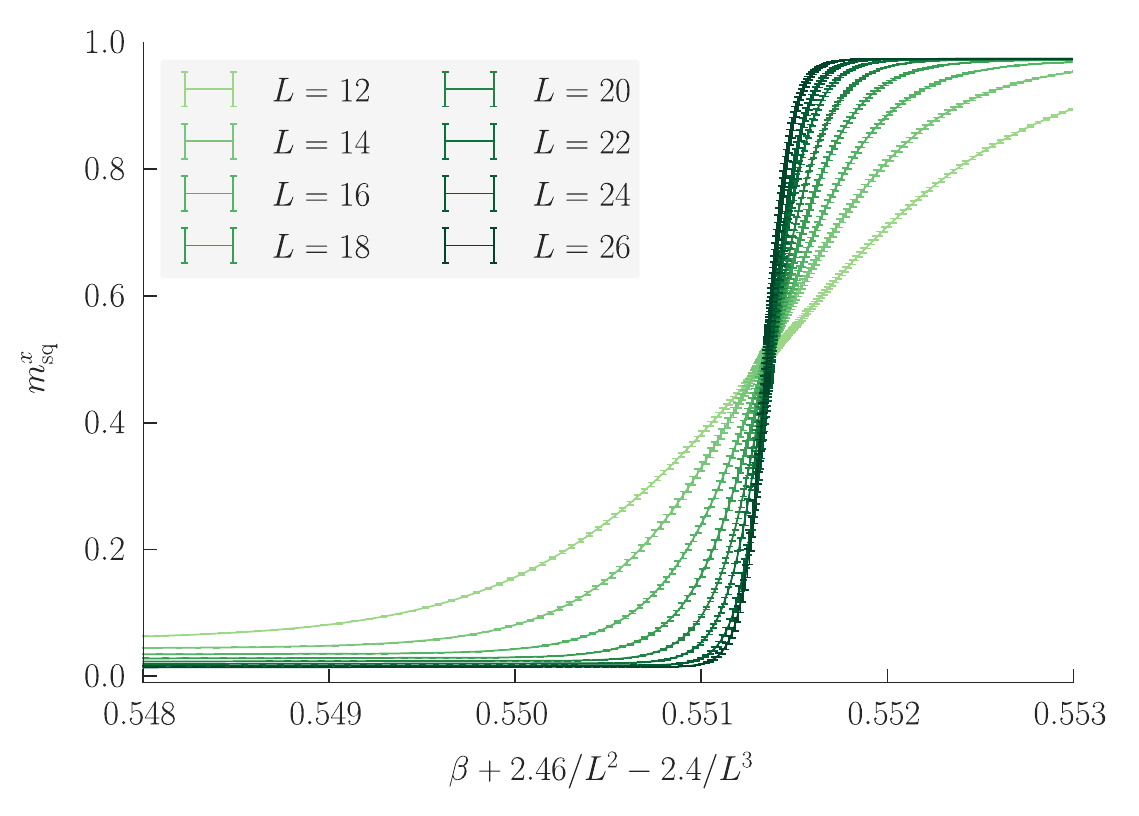}\\
    \includegraphics[width=0.5\textwidth]{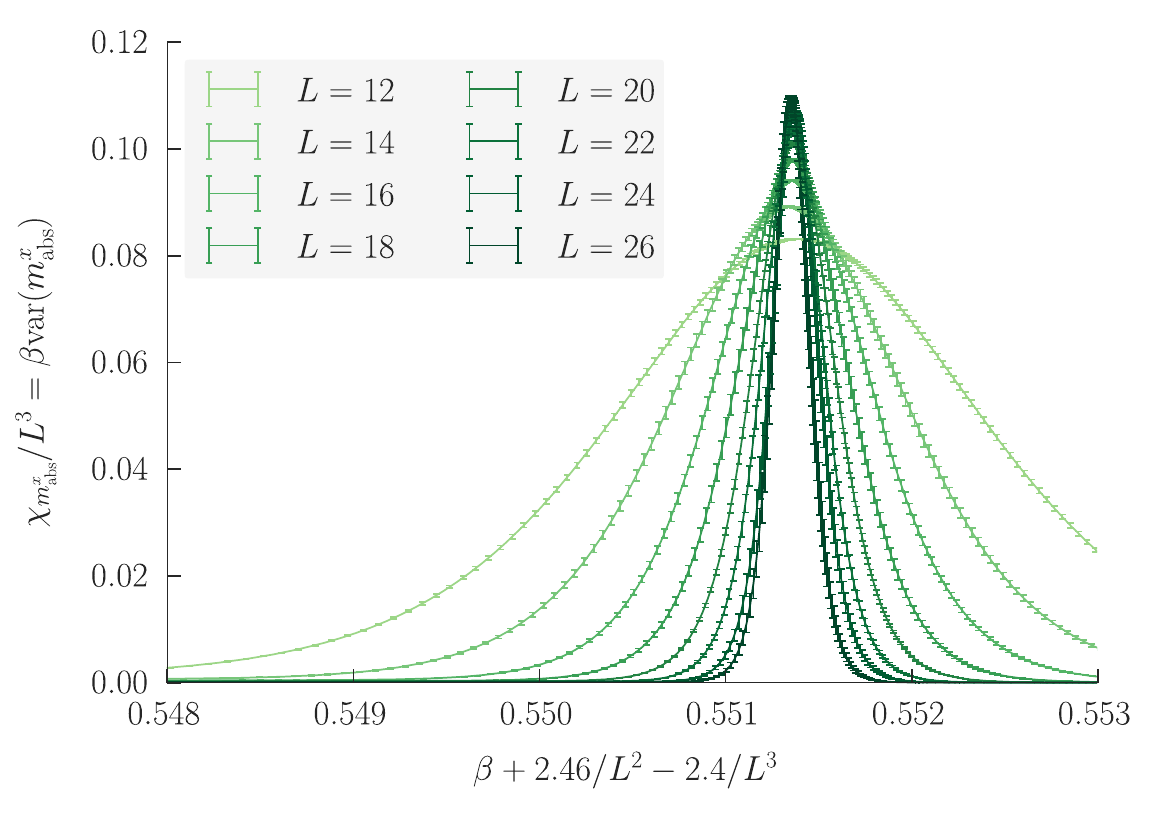}\hfill
    \includegraphics[width=0.5\textwidth]{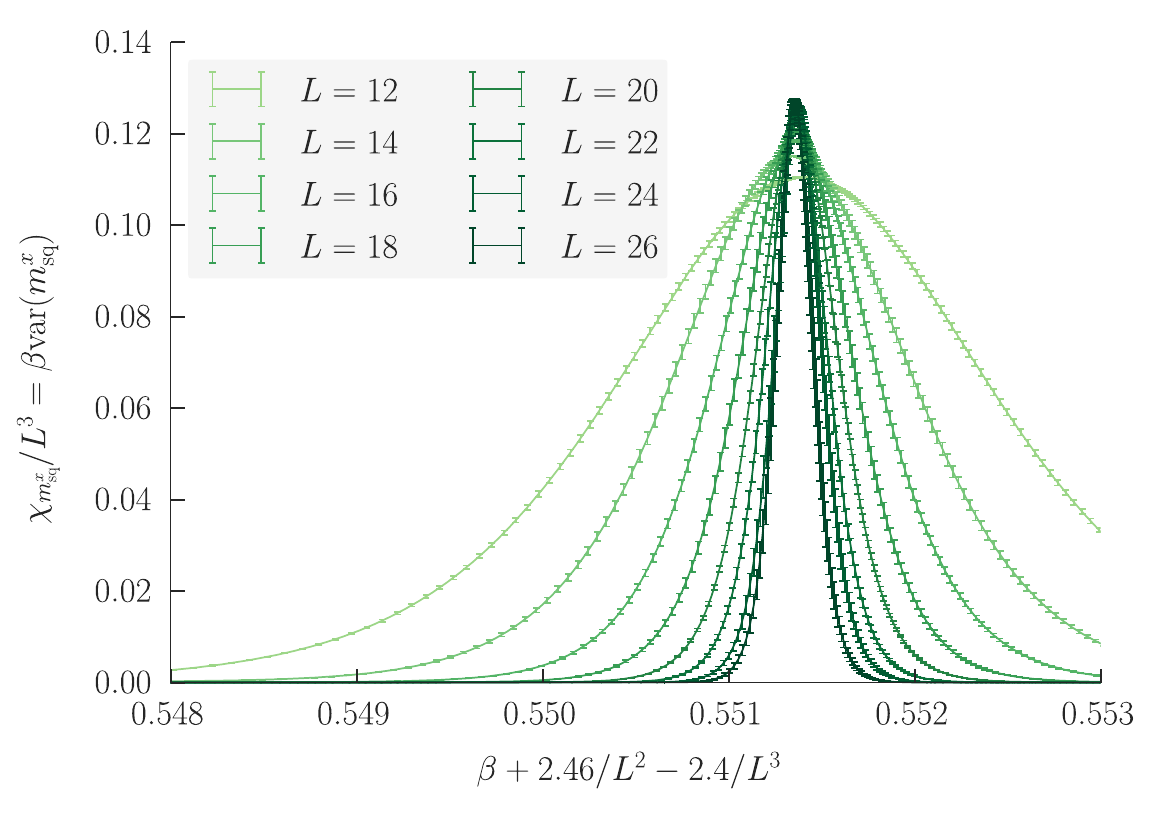}
    \caption{Canonical curves for the fuki-nuke parameters $m^{x}_{\rm abs}$
      and $m^{x}_{\rm sq}$ (upper row) along with their respective
      susceptibilities $\chi$ normalized by the system volume (lower row) over 
      shifted inverse temperature $\beta$ for
      several lattice sizes $L$.}
  \label{fig:canonmx} 
  \end{center} 
\end{figure} 

Empirically, the peak locations for the different lattice sizes $L$ can be
fitted according to the modified first-order scaling laws appropriate for
macroscopically degenerate systems discussed in detail in~\cite{goni_prl,
goni_muca},
\begin{eqnarray}
  \beta^{\chi}(L) = \beta^{\infty} + a/L^2 + b/L^3\;,
\end{eqnarray}
with free parameters $a,b$ for the available $24$ lattice sizes. Smaller
lattices are systematically omitted until a fit with quality-of-fit parameter
$Q$ bigger than $0.5$ is found. This gives for the estimate of the inverse critical
temperature $\beta^{\chi}(L)$ from the fuki-nuke susceptibility $\chi_{m_{\rm
abs}^{x}}$
\begin{eqnarray}
\label{eq:chiscaling}
  \beta^{\chi_{m_{\rm abs}^{x}}}
  (L) = 0.551\,37(3) - 2.46(3)/L^2 + 2.4(3)/L^3\;,
\end{eqnarray}
with a goodness-of-fit parameter $Q=0.64$ and $12$ degrees of freedom left.
Fits to the other directions $m_{\rm abs}^{y,z}$ and fits to the peak location
of the susceptibilities of $m_{\rm sq}^{x,y,z}$ give  the same parameters
within error bars and are of comparable quality. The estimate of the phase
transition temperature obtained here from the finite-size scaling of the
fuki-nuke order parameter(s), $\beta^\infty = 0.551\,37(3)$, is thus in good
agreement with the earlier estimate $\beta^\infty = 0.551\,334(8)$ reported
in~\cite{goni_muca} using fits to the peak location of Binder's energy
cumulant, the specific heat and the value of $\beta$ where the energy
probability density has two peaks of the same height or same weight.
Interestingly, we find that the value of the coefficient for the leading
correction also coincides.  Assuming that  the coefficient $a=-2.46(3)$ of the
leading correction is related to the inverse latent heat by  $ a =
-3 \ln(2)/\Delta\hat e$, as with the  previous estimates
\cite{goni_muca}, we find from  Eq.~(\ref{eq:chiscaling}) that $\Delta\hat e =
0.845(8)$, in good agreement with the latent heat~$\Delta\hat e=0.850\,968(18)$
reported earlier in \cite{goni_muca}.  For visual confirmation of the
finite-size scaling, the fuki-nuke magnetizations $m^x_{\rm abs, sq}$ along
with their susceptibilities divided by the system volume are plotted in
Fig.~\ref{fig:canonmx} by shifting the $x$-axis according to the scaling law,
incorporating the fit parameters. The peak locations of the susceptibilities
then all fall {on} the inverse transition temperature.
%
\begin{figure}[bt] 
  \begin{center} 
    \includegraphics[width=0.5\textwidth]{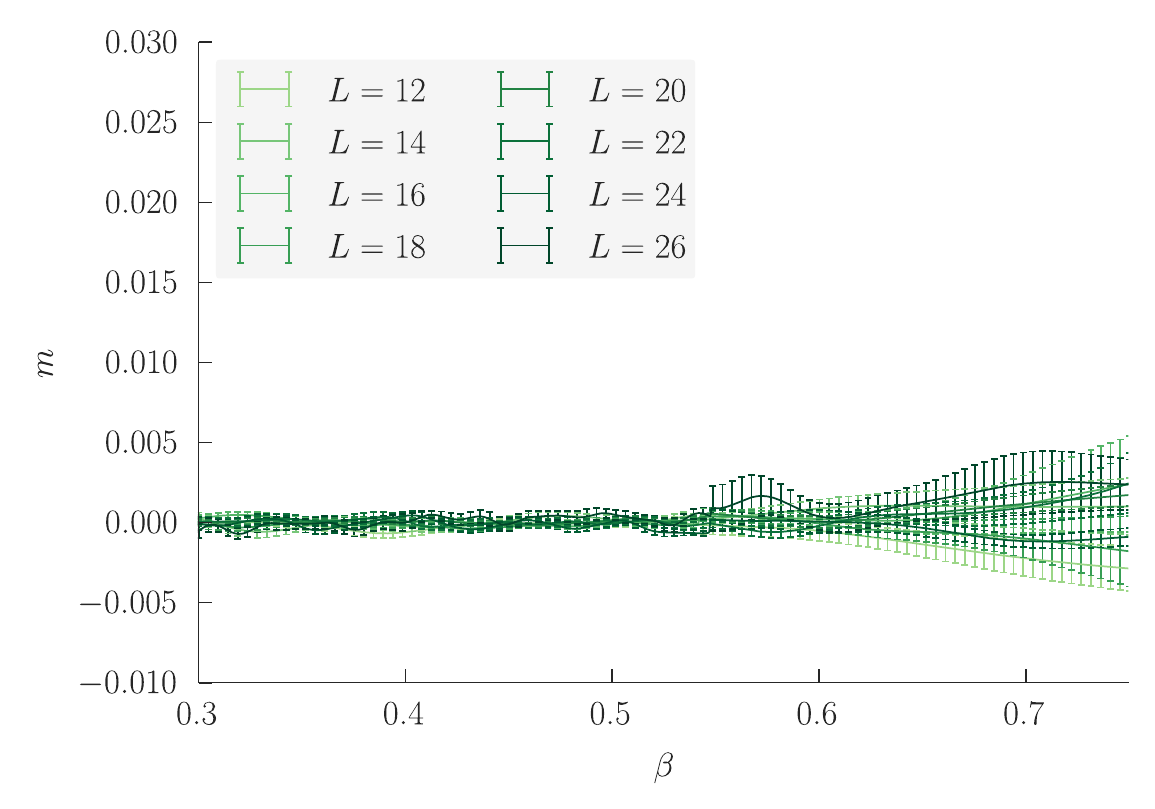}\hfill
    \includegraphics[width=0.5\textwidth]{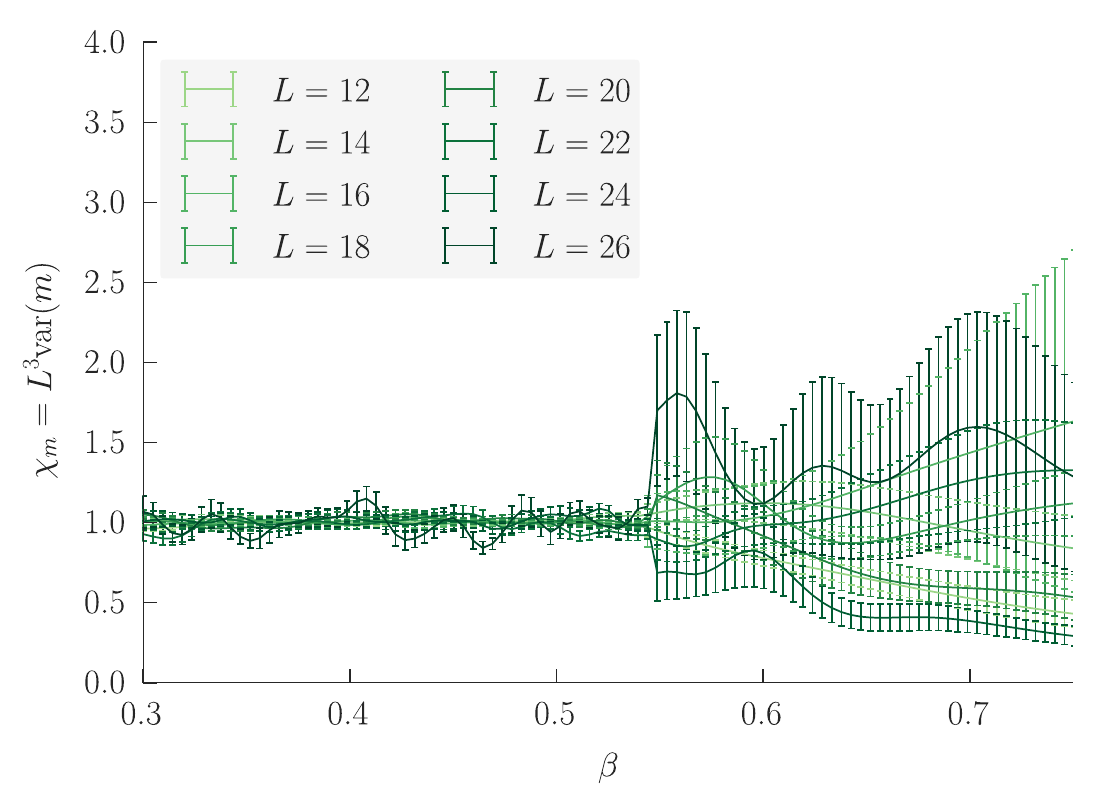}\\
    \caption{Canonical curves for the magnetization $m$ and the
      susceptibility $\chi_m$ over a broad range of inverse temperatures
      $\beta$ for several lattice sizes $L$.}
  \label{fig:canonm} 
  \end{center} 
\end{figure} 

In the earlier work~\cite{goni_order} the susceptibility {$\chi_m$} of the
standard magnetization $m$ unexpectedly behaved like an order parameter and it
continues to behave idiosyncratically in the multicanonical simulations, but in
a different manner. For compatibility with~\cite{goni_order}, the magnetic
susceptibility divided by the inverse  temperature, $\chi_m = L^3(\langle
m^2\rangle - \langle m \rangle^2)$,  is plotted in Fig.~\ref{fig:canonm} along
with the standard magnetization on a very small vertical scale (note that $m$
should be between $-1$ and $+1$). $\chi_m = 1$ in the high-temperature phase
but for the ordered, low-temperature phase the error rapidly increases below
the transition temperature, though it is clear that the susceptibility is
non-zero in this case too.  Since $\langle m \rangle = 0$, the behaviour of
$\langle m^2 \rangle$ can provide insight into this behaviour of the
susceptibility. Above the transition temperature in the high-temperature  phase
the sum over the free spin variables behaves  like a random walk with unit
step-size, therefore the expectation value of the squared total magnetization
is given by $\langle M^2 \rangle = L^3$. Taking the normalization $m = M/L^3$
into account gives $\chi_m = 1$ in this region, as seen in
Fig.~\ref{fig:canonm}. 

Below the transition temperature, it is plausible that simulations in general
get trapped in the vicinity of one of the degenerate low-temperature phases,
each of which will have a different magnetization. A canonical simulation
cannot overcome the huge barriers in the system and ``freezes'' with the same
magnetization that the system had when entering the ordered phase. This
accounts for the zero variance  seen in the Metropolis simulations
of~\cite{goni_order} below the transition temperature, since $\left< m \right>$
is frozen.  In multicanonical simulations, on the other hand, the system
travels between ordered and disordered phases, thus picking one of the possible
magnetizations each time it transits to an ordered phase  which it sticks with
until it decorrelates  again in the disordered phase. Therefore, what is seen
in the low-temperature region of Fig.~\ref{fig:canonm} for $\chi_m$ is that the
variance of $m$ is taking on rather arbitrary values due to the low statistics
obtained compared to the large number, $q=2^{3L}$, of degenerate phases one
would have to visit to sample $\langle m^2 \rangle$ properly. Even with
multicanonical simulations it is not possible to visit all of these
macroscopically degenerate phases, and the increasing error bars reflects this.
In the canonical case one gets stuck with one magnetization and would not
notice the different values, leading to much more severe ergodicity problems in
the finite simulation runs. 

\begin{figure}[t]
  \begin{center} 
    \subfigure[]{\includegraphics[width=0.5\textwidth]{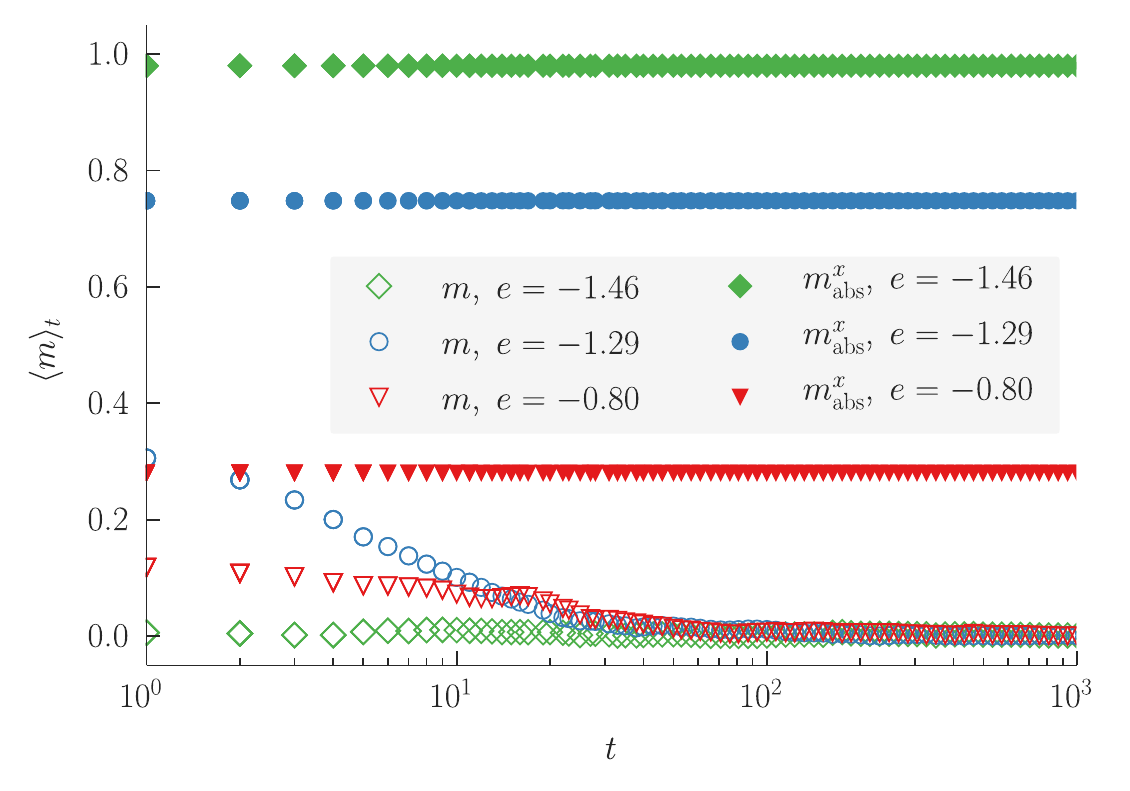}}\hfill
    \subfigure[]{\includegraphics[width=0.5\textwidth]{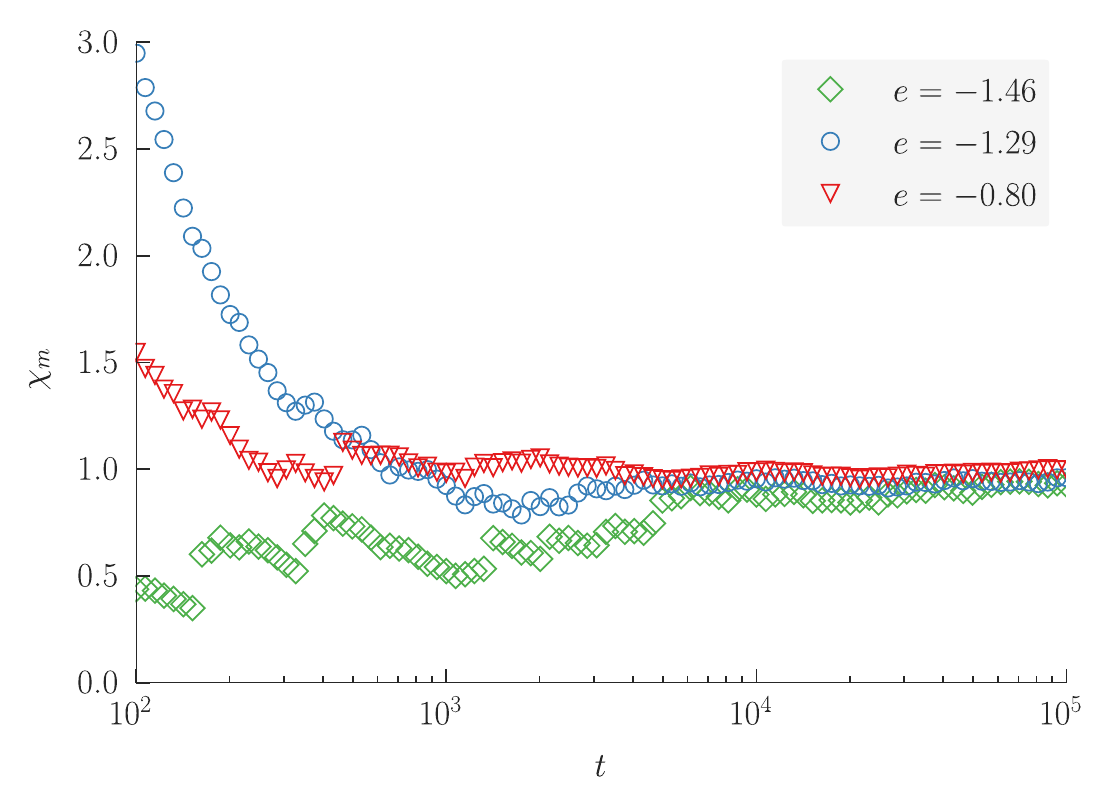}}
    \caption{(a) Running average of the standard magnetization $m$ and
      fuki-nuke parameter $m_{\rm abs}^x$ plotted against the number $t$ of
      plane-flips for three random realizations of the ordered ($e=-1.46$),
      intermediate ($e=-1.29$) and disordered ($e=-0.80$) configurations of a
      lattice with linear size $L=10$. (b) Running average of the standard magnetic
      susceptibility $\chi_m$ plotted against the number $t$ of plane-flips for the
      same three realizations. Note that the $t$-axis starts at $10^2$ because
      $\chi_m$, being  a variance, needs sufficiently many measurements to be
      meaningful.}
    \label{fig:plane-flips} 
  \end{center} 
\end{figure}

\begin{figure}[t]
  \begin{center} 
    \includegraphics[width=0.5\textwidth]{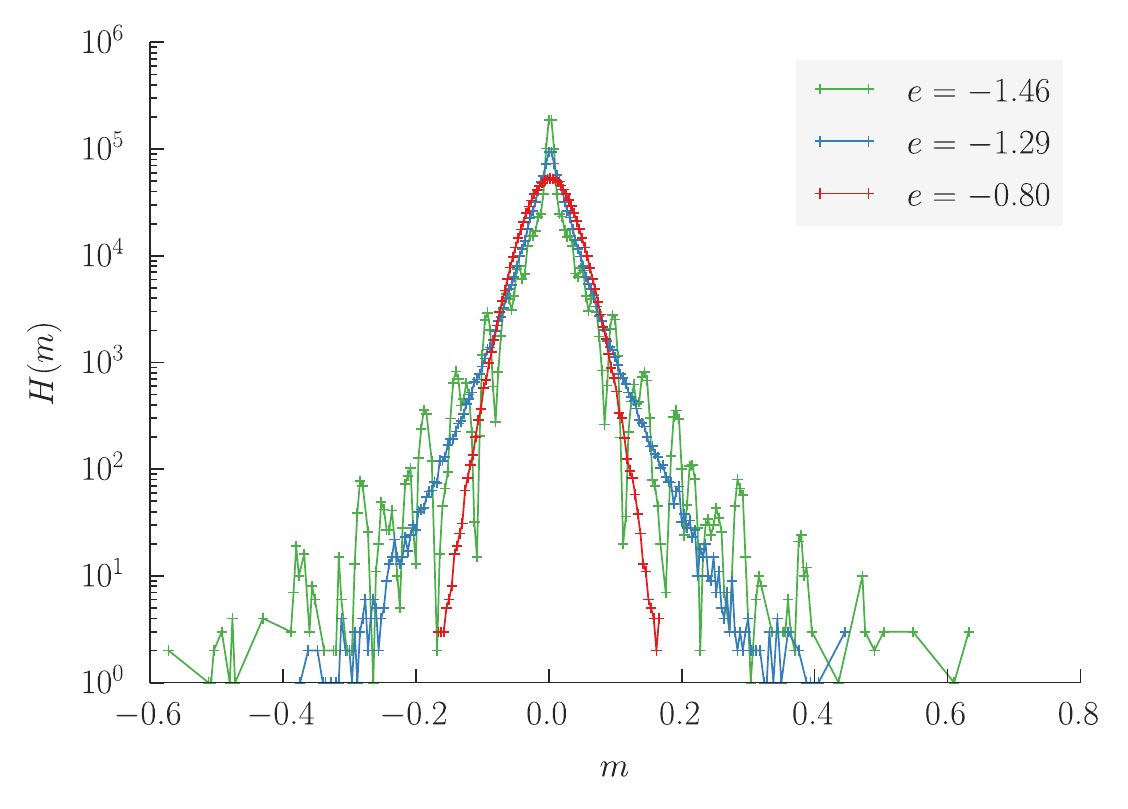}\hfill
    \caption{The histograms $H(m)$ of the standard magnetization for a total number of $10^6$ random plane-flips on a semi-logarithmic scale.}
    \label{fig:plane-flips-hist} 
  \end{center} 
\end{figure}

We investigate further the behaviour of the standard magnetization and
susceptibility and fuki-nuke magnetizations in the model by preparing several
fixed configurations with a given magnetization for a lattice with $10^3$ spins
and then \emph{only\/} flipping complete planes of spins (a ``flip-only''
update), measuring the running average of the magnetization and fuki-nuke
parameters. An example with the first thousand out of a total of $10^6$
measurements is shown in Fig.~\ref{fig:plane-flips} for three configurations
picked at random from the ordered ($e=-1.46$), intermediate ($e=-1.29$) and
disordered ($e=-0.80$) regions,
respectively, along with the histograms for the magnetization in
Fig.~\ref{fig:plane-flips-hist} obtained using the non-local flip-only update.
It can be seen that whatever the initial value the running average of the
(standard) magnetization becomes zero if one takes a long enough run so, as
expected, the flip symmetry precludes a non-zero value. The fuki-nuke order
parameters, on the other hand, should be invariant with respect to the
plane-flip symmetry and this is clearly the case in
Fig.~\ref{fig:plane-flips}(a), where the values of $m_{\rm abs}^x$ remain
constant for the ordered, intermediate and disordered starting configurations.
%

The running average of the standard magnetic susceptibility $\chi_m$  is
plotted in a similar fashion in Fig.~\ref{fig:plane-flips}(b), where it can be
seen that the ordered, intermediate  and disordered  configurations all
converge after initial transients to values of $\chi_m$ close to 1. The
non-local plane-flips thus allow enough variability in the magnetization for
the expected susceptibility value of $\chi_m=1$ to be at least approximately
attained even in the ordered phase, unlike the case of purely local  spin
flips. This suggests there might be some utility in incorporating such moves
into a Metropolis simulation of the plaquette model to improve the  ergodicity
properties.

The histograms $H(m)$ of the magnetization shown in
Fig.~\ref{fig:plane-flips-hist} on a semi-logarithmic scale display interesting
behaviour. They are symmetric around zero for all of the starting
configurations because of the $\mathbb{Z}_2$ symmetry of the Ising spins but
they have rather different shapes in each case. The disordered starting
configuration ($e=-0.80$) has a smooth maximum at $m=0$, but both the
intermediate ($e=-1.29$) and ordered ($e=-1.46$) starting configurations
generate sharp peaks at $m=0$. The pronounced peaks and valleys in the ordered
histogram are presumably a consequence of the difficulty of reaching certain
magnetization values (and the greater ease of reaching others) from an ordered
starting configuration using only plane flips. 

It would be  interesting to construct and simulate a multimagnetic
ensemble~\cite{muma}, where the weights give constant transition rates between
configurations with different magnetizations to elucidate further on the
magnetic and geometric barriers, as well as to confirm that $\langle m \rangle
= 0$ and $\chi = 1$ for the low-temperature phase.

\section{Conclusions}

The multicanonical simulations presented here provide strong support for the
idea that the plaquette  gonihedric Ising model displays the same planar,
fuki-nuke order seen in the strongly anisotropic limit of the model. In
addition, the finite-size scaling analysis of the fuki-nuke order parameters
gives scaling exponents in good agreement with the analysis of energetic
quantities carried out in \cite{goni_muca} and clearly displays the effect of
the macroscopic low-temperature phase degeneracy on the corrections to scaling.
The transition temperature  obtained here from the scaling of the fuki-nuke
order parameters, and the amplitude for the leading correction to scaling term
were found to be the same as those extracted from energetic observables. The
analysis of the magnetic order parameters carried out here is thus
complementary to the analysis of energetic observables in \cite{goni_muca} and
fully consistent with it.

The peculiar behaviour of the susceptibility of the standard magnetization
$\chi_m$ in the earlier Metropolis simulations of \cite{goni_order} was
confirmed to be an artefact of the employed algorithm.  However, even with
multicanonical simulations, sampling the macroscopically degenerate
low-temperature phase 
efficiently
is difficult. Such problems could in
principle be eased by introducing  the plane-flips which provide a valid Monte
Carlo update themselves and we investigated the behaviour of the standard
magnetization and the fuki-nuke order parameters when such updates were
applied.  

With the work reported here on magnetic observables and the earlier
multicanonical investigations of energetic quantities in \cite{goni_prl,
goni_muca} the equilibrium properties of the $3d$ plaquette gonihedric Ising
model are now under good numerical control and the order parameter has been
clearly identified. In the light of this clearer understanding, it would be
worthwhile re-investigating {\it non\/}-equilibrium properties, in particular
earlier suggestions \cite{castelnovo, glassy, Johnston2008} that the model
might serve as a generic example of glassy behaviour, even in the absence of
quenched disorder.

\section*{Acknowledgements}
This work was supported by the Deutsche Forschungsgemeinschaft (DFG)
through the Collaborative Research Centre SFB/TRR 102 (project B04) and by
the Deutsch-Franz\"osische Hochschule (DFH-UFA) under Grant No.\ CDFA-02-07.

 
\end{document}